\documentclass[aps, prl, 10pt, twocolumn, superscriptaddress, floatfix, longbibliography]{revtex4-1}
\usepackage{anyfontsize}
\usepackage{graphicx}
\usepackage{placeins} 
\usepackage{amssymb}
\usepackage{amsmath}
\usepackage{mathtools}
\usepackage{verbatim}
\usepackage{enumitem}
\usepackage[unicode=true, bookmarks=false, breaklinks=false,pdfborder={0 0 1},backref=false,colorlinks=true]{hyperref}
\hypersetup{
 linkcolor=[rgb]{0,0,1},citecolor=[rgb]{0,0,1},urlcolor=[rgb]{0,0,1}}
\mathchardef\myphen="2D
\newcommand{\citenumerics}{\cite{Qin2020-AbsenceSuperconductivityPureb, Jiang2021-GroundstatePhaseDiagram, Xu2024-CoexistenceSuperconductivityPartially, Jiang2024-GroundstatePhaseDiagram, Chen2025-GlobalPhaseDiagram, Lu2024-EmergentSuperconductivityCompeting, Xu2022-StripesSpindensityWaves, Shen2024-GroundStateElectrondopeda, Zheng2017-StripeOrderUnderdoped, Roth2025-SuperconductivityTwodimensionalHubbarda, Jiang2025-CompetitionChargedensitywaveSuperconductinga, Gu2025-SolvingHubbardModel, Liu2025-AccurateSimulationHubbarda}}
\begin{document}
\title{Large Isolated Stripes on Short 18-leg $t$-$J$ Cylinders}
\author{Tizian~Blatz}
\email{blatz.tizian@physik.uni-muenchen.de}
\affiliation{Department of Physics and Arnold Sommerfeld Center for Theoretical Physics (ASC), Ludwig-Maximilians-Universit\"at M\"unchen, Theresienstr. 37, Munich D-80333, Germany}
\affiliation{Munich Center for Quantum Science and Technology (MCQST), Schellingstr. 4, Munich D-80799, Germany}
\author{Sebastian Paeckel}
\affiliation{Department of Physics and Arnold Sommerfeld Center for Theoretical Physics (ASC), Ludwig-Maximilians-Universit\"at M\"unchen, Theresienstr. 37, Munich D-80333, Germany}
\affiliation{Munich Center for Quantum Science and Technology (MCQST), Schellingstr. 4, Munich D-80799, Germany}
\author{Ulrich~Schollwöck}
\affiliation{Department of Physics and Arnold Sommerfeld Center for Theoretical Physics (ASC), Ludwig-Maximilians-Universit\"at M\"unchen, Theresienstr. 37, Munich D-80333, Germany}
\affiliation{Munich Center for Quantum Science and Technology (MCQST), Schellingstr. 4, Munich D-80799, Germany}
\author{Fabian~Grusdt}
\affiliation{Department of Physics and Arnold Sommerfeld Center for Theoretical Physics (ASC), Ludwig-Maximilians-Universit\"at M\"unchen, Theresienstr. 37, Munich D-80333, Germany}
\affiliation{Munich Center for Quantum Science and Technology (MCQST), Schellingstr. 4, Munich D-80799, Germany}
\author{Annabelle~Bohrdt}
\email{a.bohrdt@lmu.de}
\affiliation{Department of Physics and Arnold Sommerfeld Center for Theoretical Physics (ASC), Ludwig-Maximilians-Universit\"at M\"unchen, Theresienstr. 37, Munich D-80333, Germany}
\affiliation{Munich Center for Quantum Science and Technology (MCQST), Schellingstr. 4, Munich D-80799, Germany}
\date{\today}
\begin{abstract}
Spin-charge stripes belong to the most prominent low-temperature orders besides superconductivity in high-temperature superconductors.
This phase is particularly challenging to study numerically due to finite-size effects.
By investigating the formation of long, isolated stripes, we offer a perspective complementary to typical finite-doping phase diagrams.
We use the density-matrix renormalization group algorithm to extract the ground states of an $18$-leg cylindrical strip geometry, making the diameter significantly wider than in previous works.
This approach allows us to map out the range of possible stripe filling fractions on the electron versus hole-doped side. We find good agreement with established results, suggesting that the spread of filling fractions observed in the literature is governed by the physics of a single stripe.
Taking a microscopic look at stripe formation, we reveal two separate regimes -- a high-filling regime captured by a simplified squeezed-space model and a low-filling regime characterized by the structure of individual pairs of dopants.
Thereby, we trace back the phenomenology of the striped phase to its microscopic constituents and highlight the different challenges for observing the two regimes in quantum simulation experiments.
\end{abstract}
\maketitle
\begin{figure}[t!!]
\centering
\includegraphics[width=\linewidth]{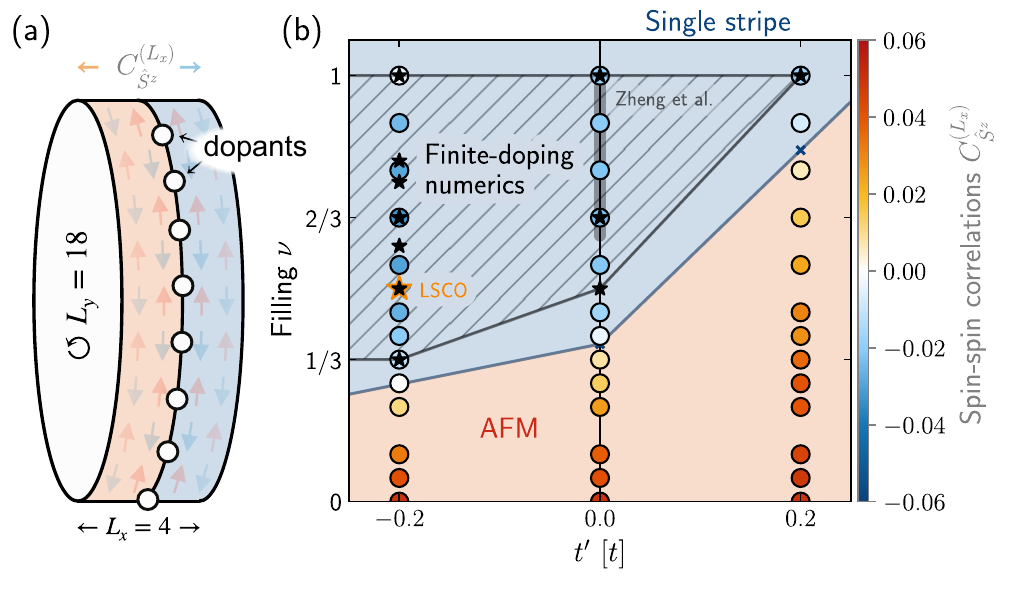}
\caption{
{Stripe Formation:}~(a) Sketch of the cylindrical system geometry. A stripe is identified with the formation of a spin domain wall across the system, indicated by the sign of the cross-lattice spin correlations $C^{(L_x)}_{\hat{S}^z}$.
(b) $t' \: - \: \nu$ diagram of stripe formation. Individual ground-state calculations are marked by points, and color-coded by $C^{(L_x)}_{\hat{S}^z}$.
On the electron-doped side ($t'>0$), we find a narrow filling region featuring negative correlations centered around $\nu=1$.
At lower values of $t'$, the domain where a stripe is formed extends to significantly lower fillings -- down to $\nu \sim 1/3$ on the hole-doped side ($t'<0$).
For comparison, the black stars mark stripe fillings stabilized
in the numerical literature{~\citenumerics} (see the Appendix
for details).
The grey shaded line at $t'=0$ indicates the near-degenerate filling fractions determined by B.-X. Zheng et al.~\cite{Zheng2017-StripeOrderUnderdoped}.
As a guide to the eye, the black hatched region visualizes the range of stabilized fillings, which is well-reflected by the single-stripe results.
In addition, the orange star indicates the stripe filling $\nu \approx 1/2$ observed in the LSCO cuprate compounds~\cite{Cheong1991-IncommensurateMagneticFluctuations, Tranquada1995-EvidenceStripeCorrelations}.
}
\label{fig_filling_t_prime_diagram}
\end{figure}

%
\textit{Introduction}---{}Many unconventional superconductors, like the cuprates, feature density-wave order in close proximity to superconductivity~\cite{Tranquada1995-EvidenceStripeCorrelations, vonZimmermann1997-XrayScatteringStudy, Emery1999-StripePhasesHightemperaturea, Scalapino2012-CommonThreadPairing, Keimer2015-QuantumMatterHightemperature}.
To characterize these phases, a large body of numerical works{~\citenumerics} studies the ground states of the square-lattice Fermi-Hubbard~\cite{Hubbard1963-ElectronCorrelationsNarrowb} and $t$-$J$~\cite{Chao1977-KineticExchangeInteraction, Chao1978-CanonicalPerturbationExpansion, Hirsch1985-AttractiveInteractionPairing} models -- the microscopic models believed to capture the essence of cuprate physics.
When doped away from half filling, these models commonly feature a striped phase, where peaks in the charge density separate domains of opposite antiferromagnetic (AFM) order.
The interplay of this striped phase with superconductivity was at the center of early theories of unconventional superconductivity~\cite{Emery1997-SpingapProximityEffect, Kivelson1998-ElectronicLiquidcrystalPhasesa, Emery1999-StripePhasesHightemperaturea} and is the subject of ongoing study to this day.
In particular, the suggested coexistence of superconductivity with partially filled stripes~\cite{Xu2024-CoexistenceSuperconductivityPartially, Roth2025-SuperconductivityTwodimensionalHubbarda} raises the question how stripe order is connected to the models' microscopic pairing properties. \\
\indent However, the small system sizes available to even the most advanced numerical studies
constitute a serious challenge for simulating density-wave order:
The stripes formed in these systems are typically very short, making it hard to separate pairing from stripe physics~\cite{Chung2020-PlaquetteOrdinary$d$wave, Blatz2025-TwoDopantOriginCompetinga}, and coarsely discretizing the available filling fractions.
Despite this, the differences in ground-state energy between stripes of different filling can be extremely small~\cite{Qin2020-AbsenceSuperconductivityPureb, Zheng2017-StripeOrderUnderdoped}, making convergence difficult and introducing a strong dependence on additional finite-size quantities like the number of dopants or the system size parallel to the stripe-wavevector~\cite{Xu2022-StripesSpindensityWaves}.
To address these challenges, extensive works have appeared in recent years that approach the thermodynamic limit by careful finite-size extrapolation and multi-method comparison~\cite{Zheng2017-StripeOrderUnderdoped, Xu2022-StripesSpindensityWaves, Xu2024-CoexistenceSuperconductivityPartially},
or unlock unprecedented system sizes using variational methods~\cite{Liu2025-AccurateSimulationHubbarda, Gu2025-SolvingHubbardModel, Roth2025-SuperconductivityTwodimensionalHubbarda}.
However, due to the high computational costs and commensurability effects, it remains difficult to cover the wide range of dopings and model parameters needed to formulate a unified picture of the striped phase. This calls for complementary approaches and for reaching a better microscopic understanding.
\\
\indent Here, we study how dopants organize into long individual stripes in the $t$-$t'$-$J$ model.
Using the density-matrix renormalization group (DMRG) algorithm in the framework of matrix-product states (MPS), we compute the ground states of short $18$-leg cylinders, see Fig.~\ref{fig_filling_t_prime_diagram}(a).
Computations in this unusual geometry are enabled by a tailored MPS mapping that shifts the exponential computational cost away from the periodic direction.
Reducing the system to a single stripe allows us to accurately map out the region of the phase diagram that supports stripe formation without the proximity to other phases and prominent finite-size and commensurability effects that trouble other numerical works.
We investigate the microscopic mechanisms of stripe formation by probing the stripe's charge and spin structure, and by comparing the ground states' spin correlations to the predictions of a simple squeezed-space model~\cite{Ogata1990-BetheansatzWaveFunction, Schlomer2023-RobustStripesMixeddimensional}.
In doing so, we separate a low-filling regime, present only towards the hole-doped side, from a more robust high-filling regime.
We argue that the pattern in which partially filled stripes form follows from the structure of a single pair of dopants~\cite{Blatz2025-TwoDopantOriginCompetinga} -- the principal building block of both the striped and superconducting many-body phases.
\\
\indent \textit{Model and method}---{}We study the formation of individual stripes in the $t$-$t'$-$J$ model
\begin{align} \label{eq:t-J}
    &\hat{H}_{t \myphen t' \myphen J} \ = \
    -  t \sum_{\langle i, j \rangle, \sigma} (\Tilde{c}^\dagger_{i, \sigma} \Tilde{c}_{j, \sigma} + \mathrm{H.c.}) \\
    & \quad - t' \sum_{\mathclap{\langle\langle i, j \rangle\rangle, \sigma}} (\Tilde{c}^\dagger_{i, \sigma} \Tilde{c}_{j, \sigma} + \mathrm{H.c.})
    + J \sum_{\mathclap{\langle i, j \rangle}} \left( \hat{\mathbf S}_{i} \cdot \hat{\mathbf S}_{j}
    - \frac{\Tilde{n}_i \Tilde{n}_j}{4} \right) \ , \notag
\end{align}
characterized by the nearest-neighbor tunneling $t$, the diagonal next-nearest-neighbor tunneling $t'$, and the Heisenberg interaction $J$.
Here, $\Tilde{c}^\dagger_{i, \sigma} = \hat{c}^\dagger_{i, \sigma} ( 1 - \hat{n}_{i, - \sigma})$ creates a particle with spin $\sigma$, but disallows double occupancies.
The operator is defined via the usual fermionic annihilation (creation) operator $\hat{c}^\dagger_{i, \sigma}$ at coordinate $i = (x, y)$ and the number operator
$\hat{n}_{i, \sigma}$. \\
The $t$-$t'$-$J$ model is a popular microscopic model to describe the cuprate superconductors~\cite{Zhang1988-EffectiveHamiltonianSuperconducting}.
It is connected to the paradigmatic Fermi-Hubbard model by the Schrieffer-Wolff transformation~\cite{Auerbach1994-InteractingElectronsQuantuma}.
In our work, we fix $t = 1$ as the unit of energy and investigate the model's properties at $J = 1/2$ and $t' \in [ 0 ; \: \pm 0.2 ]$.
This value of $J$ is realistic for the study of cuprate materials~\cite{Hybertsen1989-CalculationCoulombinteractionParameters, Dagotto1994-CorrelatedElectronsHightemperature}.
As has been demonstrated in large-scale numerical simulations, the next-nearest-neighbor tunneling $t'$-term has important implications for superconductivity and stripe formation in the system's ground state.
In the cuprates, band-structure calculations estimate the strength of this term to be $t' \approx -0.2 \: t$~\cite{Andersen1995-LDAEnergyBands, Hirayama2018-InitioEffectiveHamiltonians}. Flipping the sign of $t'$ is equivalent to exchanging electron and hole dopants by means of a particle-hole transformation, which allows us to study both sides of the doping diagram.
\\
\indent In the following, we study the formation of single stripes in the ground states of wide cylindrical systems using DMRG~\cite{White1992-DensityMatrixFormulationb, White1993-DensitymatrixAlgorithmsQuantuma, Schollwoeck2011-DensitymatrixRenormalizationGroup}.
As is typical for DMRG studies, the system features periodic boundary conditions (PBCs) in the $y$-direction and open boundary conditions (OBCs) in the $x$-direction.
However, to study long isolated stripes, we chose the non-standard, high aspect ratio $L_y > L_x$.
For the most part, we study systems of $L_x \times L_y = 4 \times 18$ sites, i.e. $18$-leg cylinders.
This cylinder width is significantly larger than the typical $L_y$ studied in finite-doping phase diagrams~\cite{Jiang2021-GroundstatePhaseDiagram, Shen2024-GroundStateElectrondopeda, Jiang2024-GroundstatePhaseDiagram, Lu2024-EmergentSuperconductivityCompeting, Jiang2025-CompetitionChargedensitywaveSuperconductinga}, which use low aspect ratios $L_y / L_x \lesssim 1$.
We apply weak chemical potential offsets $\Delta \mu = 3/4 \: J$ on the edges (i.e., $x \in [1, L_x]$) to lift a possible competition between vertical and horizontal stripes in the system.
To enable ground-state searches in this geometry, we use a tailored MPS mapping, inspired by one-dimensional systems with PBCs~\cite{Wilke2023-SymmetryprotectedBoseEinsteinCondensation} (details in the Appendix).
In essence, this mapping shifts the exponential complexity in the periodic direction $L_y$ -- the usual computational bottleneck -- to exponential complexity in $2 \: L_x$.
Therefore, the $L_x = 4$ systems we study are comparable in computational effort to $L_y = 8$ cylinders with conventional MPS mappings.
\\
\indent \textit{Stripe formation}---{}The primary parameter of the striped phase is the stripe filling fraction $\nu = N^\mathrm{h} / L_y$, where $N^\mathrm{h}$ is the number of dopants in a stripe. In an extended system at finite doping $\delta$, this sets the stripe wavelength $\lambda = \nu / \delta$, in units of the lattice constant ($a = 1$).
Large-scale numerical studies of the ground state phase diagram have revealed a large number of stabilized filling fractions, depending on the system's parameters, size, and boundary conditions.
These range from fully filled stripes ($\nu = 1$) to fillings as low as $\nu = 1/3$.
Even though the typically small extent $L_y$ heavily discretizes the filling fractions available,
the differences in energy between different fillings can be extremely small~\cite{Qin2020-AbsenceSuperconductivityPureb, Zheng2017-StripeOrderUnderdoped, Xu2022-StripesSpindensityWaves}.
Therefore, an important task is to determine the filling region that supports the formation of stripes in a manner that is robust to finite-size effects.
This question can be addressed by studying the formation of a single stripe.\\
\indent At a given number of dopants in the system, we identify a stripe via the staggered cross-lattice spin correlations
\begin{equation}
    C^{(L_x)}_{\hat{S}^z} = \frac{1}{L_y} (-1)^{L_x-1} \: \sum_{y=1}^{L_y} \langle \hat{S}^z_{(1, \: y)} \hat{S}^z_{(L_x, \: y)} \rangle \ .
\end{equation}
In our geometry, positive correlations suggest uninterrupted AFM order, while negative correlations signal the presence of an AFM spin domain wall, which we identify with the formation of a stripe. \\
\indent Based on this criterion, Fig.~\ref{fig_filling_t_prime_diagram}(b) shows the filling region supporting the formation of a stripe depending on the next-nearest-neighbor tunneling $t'$.
In agreement with established results{~\citenumerics}, we find a large filling region $1/3 \lesssim \nu \lesssim 1$ featuring stripe formation on the hole-doped side ($t' = -0.2 \: t$). On the electron-doped side ($t' = 0.2 \: t$), a stripe may only be formed in a narrow region around $\nu = 1$ while uninterrupted AFM order persists to significantly larger fillings than on the hole-doped side. \\
\indent As our work approaches the striped phase from a very different perspective compared to typical finite-doping studies featuring multiple short stripes, the agreement we find is highly non-trivial.
It confirms that valuable insights about the physics on the electron versus hole-doped side can be gained by studying a single stripe.
Furthermore, it suggests that the coupling between stripes takes a secondary role compared to the formation of a single stripe in the background AFM.
This interpretation is in line with early arguments for a proximity effect~\cite{Emery1997-SpingapProximityEffect, Emery1999-StripePhasesHightemperaturea} and recent results, reporting a high compressibility of the stripe period~\cite{Zheng2017-StripeOrderUnderdoped, Xu2022-StripesSpindensityWaves}.
In addition, breaking down the many-body stripe phase to the single-stripe level allows us to relate its microscopic origin to the model's pairing properties in the following.
\\
\indent \textit{Microscopic structure}---{}Another important aspect of the striped phase concerns the stripes' microscopic structure.
\begin{figure}[t!!]
\centering
\includegraphics[width=\linewidth]{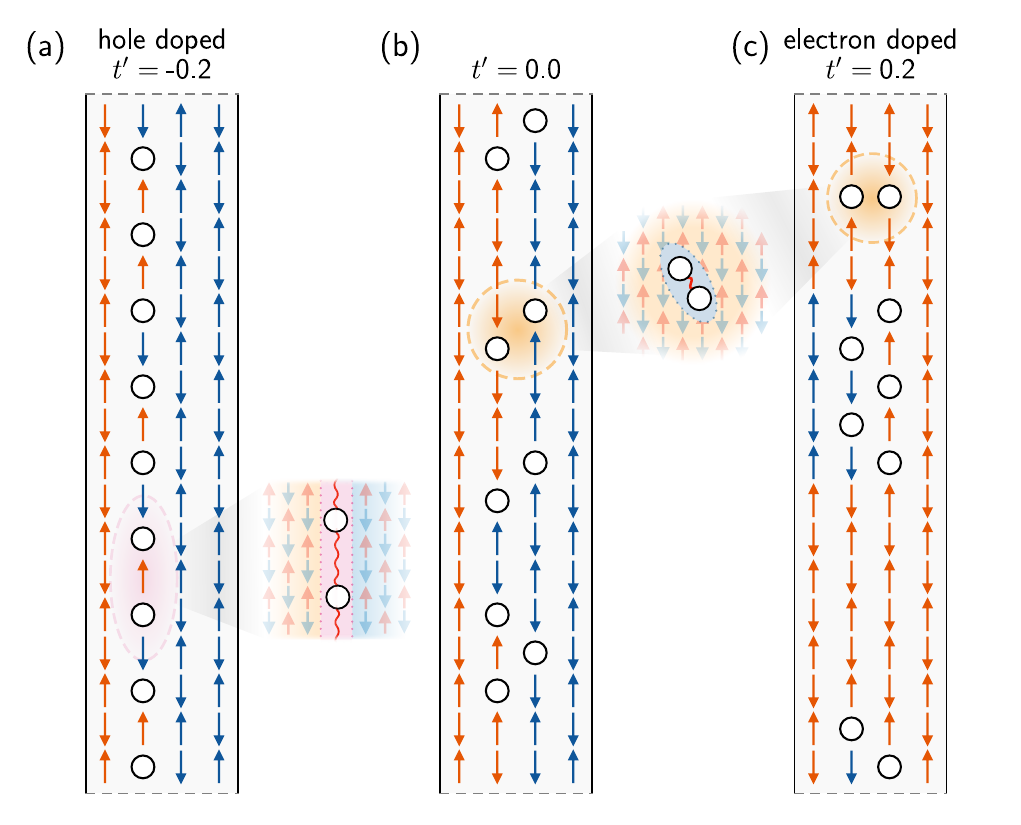}
\caption{
{Microscopic Structure:}~High-probability product states revealing the charge and spin structure of the $\nu = 1/2$ states on $4 \times 18$ cylindrical strips at different values of $t'$.
The spins are colored according to the staggered magnetization in order to visualize the formation of a spin domain wall.
(a) At $t' = -0.2$, the doped holes are mostly separate and form a spin domain wall, which we associate with the stripe state.
(b) At $t' = 0.0$, the dopants form a stripe with an internal pairing structure: A spin domain wall is formed, but the dopants combine into tightly bound pairs.
(c) At $t' = 0.2$, there is no stripe at $\nu = 1/2$. The dopants form tightly bound pairs in an uninterrupted AFM background.
}
\label{fig_internal_structure}
\end{figure}
Several studies have reported the formation of integer pair stripes (IPS), i.e., stripes that are formed out of an even number of dopants~\cite{Jiang2021-GroundstatePhaseDiagram, Qin2020-AbsenceSuperconductivityPureb, Jiang2024-GroundstatePhaseDiagram}. Their existence suggests an internal pairing structure of the stripes, and it has been argued that pairs coherently tunneling into and out of stripes constitute an important aspect of the coexistence of stripes and superconductivity~\cite{Emery1997-SpingapProximityEffect, Emery1999-StripePhasesHightemperaturea, Xu2024-CoexistenceSuperconductivityPartially}.
In contrast, stripe order in which the number of dopants per stripe is not even has recently been observed on the hole-doped side.
These so-called non-integer pair stripes (nIPS) have highlighted the possibility of striped states forming out of separate dopants~\cite{Xu2024-CoexistenceSuperconductivityPartially}. \\
\indent To uncover the microscopic structure of the extended isolated stripes, we employ high-probability product states introduced in Ref.~\cite{Jiang2021-GroundstatePhaseDiagram}.
Starting from the full MPS representing the ground state, a projective measurement is performed to fix a dopant to the site with the highest dopant density. The densities are then re-evaluated in the projected state and the procedure continues until all dopants are fixed. The spin structure is subsequently obtained in a similar manner (see~\cite{Jiang2021-GroundstatePhaseDiagram} for details) to produce a product state with high probability.
The procedure allows a glimpse into spin and charge structure without the fluctuations inherent to regular snapshots. For example, it will always produce an uninterrupted checkerboard pattern for the undoped AFM. \\
\indent Fig.~\ref{fig_internal_structure} depicts the uncovered structure at fixed $\nu = 1/2$ and varying $t'$.
At $t' \le 0$, we find striped states featuring a spin domain wall.
Importantly, we find the stripes to contain important contributions from next-nearest neighbor pairs at $t'=0$, while the dopants are separate at $t'=-0.2$.
These findings are consistent with the internal pair structure of IPS predicted in the literature on the vanilla models, which is in contrast to the possibility of nIPS reported on the hole-doped side.
Finally, on the electron-doped side ($t'=0.2$), the dopants form spatially tightly bound pairs which do not induce a spin domain wall, matching the absence of stripes reported in the literature in this regime~\cite{Xu2024-CoexistenceSuperconductivityPartially, Jiang2024-GroundstatePhaseDiagram, Lu2024-EmergentSuperconductivityCompeting}. \\
\indent All three cases relate back to the model's pairing properties reported by a previous study~\cite{Blatz2025-TwoDopantOriginCompetinga}.
On the single-pair level, the formation of tightly-bound pairs was found to be favored on the electron-doped side, while the hole-doped side features extended pairs associated with the formation of a spin-domain wall.
Extending the cylinder diameter now allows us to observe how multiple of these pairs either form a collective stripe instability, or coexist in an uninterrupted AFM background.
Thereby, we conclude a second important step toward understanding the striped phase through its microscopic constituents.
In the first step, presented in the previous section, we were able to explain the filling fractions reported in the macroscopic stripe phase through the properties of a single stripe.
In addition, here we demonstrate that the formation of these single stripes is shaped by the real-space structure of individual pairs of dopants.
\\
\indent \textit{Squeezed space}---{}Finally, we take a more quantitative look at the cross-lattice spin correlations $C_{\hat{S}^z}^{(L_x)}$
and identify two distinct regimes of stripe formation.
\begin{figure}[t!!]
\centering
\includegraphics[width=\linewidth]{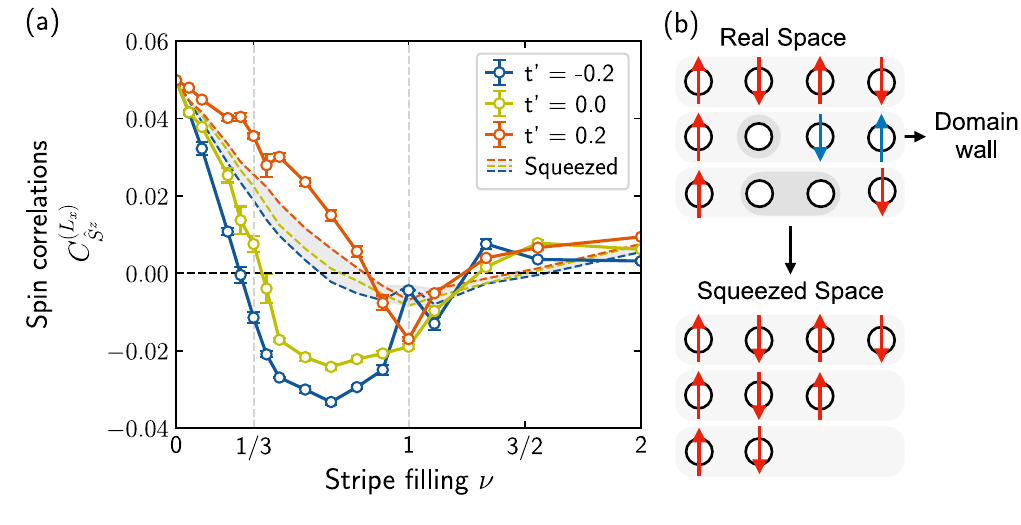}
\caption{
{Spin Correlations:}~(a)
Cross-lattice spin correlations $C^{(L_x)}_{\hat{S}^z}$ averaged over all $1 \leq y \leq L_y$.
Negative correlations indicate the formation of a spin domain wall.
The lines show the correlations in the ground state of an $4 \times 18$ system at different values of $t'$, scanning the stripe filling fraction $\nu$.
The dashed lines show the predictions of a simplified squeezed-space model.
(b) Squeezed space picture for the formation of a spin domain wall: 
Focussing on a fixed-$y$ cut through the system, a single hole is expected to lead to a spin-domain wall (center) while an even number of holes does not disturb the AFM (top, bottom).
}
\label{fig_szszs}
\end{figure}
Fig.~\ref{fig_szszs}(a) shows a line plot of the data color-coded in Fig.~\ref{fig_filling_t_prime_diagram}.
The zero crossings mark the boundaries of the stripe-stability regions.
The quantitative comparison of the different values of $t'$ reveals a strong $t'$-dependence at $0 < \nu < 1$, contrasted by a very weak dependence at $\nu = 0$ and $\nu \ge 1$.
As is explained below, we can understand this feature based on the results of the previous section, combined with a simplified squeezed space picture. \\
\indent In the $0 < \nu < 1$ region, 
the previous section uncovered
a distinct charge and spin structure for each value of $t'$.
Tightly bound pairs on the electron-doped side exist in an uninterrupted AFM background, translating to positive correlations.
At $t' \le 0$, extended distance-$3$ pairs~\cite{Blatz2025-TwoDopantOriginCompetinga} allow the formation of stripes at fillings as low as $\nu = 1/3$.
Thus, we identify the origin of the strong $t'$-dependence at low fillings to be the different pairing properties on the electron-doped versus hole-doped side. \\
At high filling $\nu \sim 1$, we can think of the stripe as a fully filled, one-dimensional line of holes without any additional pairing structure, which can explain the emergence of a spin domain wall by a simple squeezed space picture comparable to spin-charge separation in one dimension:
As sketched in Fig.~\ref{fig_szszs}(b) for a cut of the lattice at fixed $y$, we can think of the dopants displacing the Néel spin background.
In a chain with a single hole, this introduces a spin domain wall, in agreement with the $\nu = 1$ observation of Fig.~\ref{fig_szszs}(a), independent of $t'$. \\
\indent We can quantify this squeezed space prediction by replacing the Néel background by the true strength of the spin-spin correlations in the undoped Heisenberg AFM, $\langle \hat{S}^z_0 \hat{S}^z_x \rangle_{\nu=0}$.
Via the hole densities and two to four-point hole-hole correlation functions, we can access the probability of each charge configuration in a single-$y$ cut of the lattice.
This allows constructing a filling-dependent squeezed-space prediction for the cross-lattice spin-spin correlations by combining the charge information with the spin-correlations of the undoped AFM (details in the Appendix).
In Fig.~\ref{fig_szszs}(a) this prediction is compared to the true spin correlations at finite filling.
As expected, it captures the formation of a domain wall around $\nu = 1$, as well as the insensitivity to $t'$ and vanishing of the domain wall in the high-filling region $1 < \nu \leq 2$.
However, the squeezed-space picture cannot explain the behavior and strong $t'$-dependence we observe in the low-filling region -- consistent with our interpretation that stripe formation is driven by pairing in this regime.
\\
\indent \textit{Discussion}---{}Using a tailored MPS mapping, we investigate the formation of isolated stripes in the $t$-$t'$-$J$ model, which are significantly longer than those typically accessible in the numerical study of the model's phase diagram.
This methodological advancement allows us to accurately map out the filling regions where we expect stripes to form, depending on the value of the next-nearest-neighbor tunneling $t'$.
In agreement with the extensive literature for both the $t$-$t'$-$J$ and Fermi-Hubbard models, our single-stripe perspective reveals that stripes can be formed in a large range of filling $1/3 \lesssim \nu \lesssim 1$ on the hole-doped side, while the electron-doped side is dominated by local pairing and may feature spin-charge stripes only in a narrow range around $\nu = 1$.
Thereby, we establish the microscopic origin of the range of known filling fractions using an approach that is robust to finite-size effects.
In particular, we understand the formation of partially filled stripes toward the hole-doped side through a change in the model's microscopic pairing properties.
We argue that recently observed non-integer pair stripes (nIPS) can be understood as a collective instability of loosely bound pairs.
On the flipside, we demonstrate how the formation of filled stripes can be accounted for by a simple squeezed-space picture, independent of the next-nearest-neighbor tunneling $t'$.
By combining these two observations, our work provides a unified perspective for understanding the coexistence of pairing with partially filled stripes, in contrast to the insulating nature of the filled-stripe phase~\cite{Zheng2017-StripeOrderUnderdoped, Qin2020-AbsenceSuperconductivityPureb, Xu2024-CoexistenceSuperconductivityPartially}.
\\
\indent Our work highlights the importance of microscopic building blocks for understanding the complex many-body phases and presents a step towards unifying the diverse numerical literature about the striped phase.
Additionally, we expect the two filling regimes we identify to be consequential for the realization of striped phases in quantum simulation platforms.
The squeezed-space picture is highly robust and extends to the mixed-dimensional setting~\cite{Schlomer2023-RobustStripesMixeddimensional}, where the formation of a filled stripe has recently been observed in an ultracold-atom experiment~\cite{Bourgund2025-FormationIndividualStripes}.
Meanwhile, we directly relate stripes of lower filling to pairing, meaning that a realization will likely require temperatures on the order of the binding energy $0.1 \: t$ to $0.2 \: t$~\cite{Blatz2025-TwoDopantOriginCompetinga}.
As quantum simulation platforms approach these low temperatures~\cite{Xu2025-NeutralatomHubbardQuantuma}, building a microscopic understanding of the processes involved becomes increasingly important.
\\
\\
\indent \textit{Acknowledgments}---{}We are very grateful to 
Alex Deters,
Youqi Gang,
Markus Greiner,
Daniel Harrington,
Anant Kale,
Lev Kendrick,
Henning Schlömer,
Krzysztof Wohlfeld,
and
Aaron Young
for fruitful discussions and insights.
This research was funded by the Deutsche Forschungsgemeinschaft (DFG, German Research Foundation) under Germany’s Excellence Strategy—EXC-2111—390814868
and by the European Research Council (ERC) under the European Union’s Horizon 2020 research and innovation programme (grant agreement number 948141).
\\
\\
\indent \textit{Data availability}---{}The data presented in the figures of the main text is available at \url{https://github.com/TizianBlatz/stripes_tJ}.
\FloatBarrier
\bibliography{stripe_paper, suppmat}
\clearpage
\appendix
\section{Literature Reference}
Here, we provide a detailed overview of the stripe filling fractions reported in numerical ground-state works in the Fermi-Hubbard and $t$-$t'$-$J$ models, used for the literature reference presented as part of Fig.~\ref{fig_filling_t_prime_diagram}.
In summary, the reported stabilized stripe fillings for either model, depending on the value of $t'$, are
\begin{align}
    \nu_{t'=-0.2} &\in [1/3, \: 1/2, \: 3/5, \: 2/3, \: 3/4, \: 4/5, \: 1] \\
    &\quad \ \text{(see Refs.~\cite{Xu2024-CoexistenceSuperconductivityPartially, Jiang2024-GroundstatePhaseDiagram, Shen2024-GroundStateElectrondopeda, Jiang2021-GroundstatePhaseDiagram, Lu2024-EmergentSuperconductivityCompeting})} \ , \notag \\
    \nu_{t'=0.0} &\in [1/2, \: 2/3, \: 1] \ \text{\cite{Qin2020-AbsenceSuperconductivityPureb, Jiang2024-GroundstatePhaseDiagram, Shen2024-GroundStateElectrondopeda, Lu2024-EmergentSuperconductivityCompeting, Xu2022-StripesSpindensityWaves, Zheng2017-StripeOrderUnderdoped}} \ , \\
    \nu_{t'=0.2} &\in [1] \ \text{\cite{Jiang2021-GroundstatePhaseDiagram}} \ .
\end{align}
On the hole-doped side, \cite{Xu2024-CoexistenceSuperconductivityPartially} presents a variety of fillings $\nu \in [3/5, \: 2/3, \: 3/4, \: 4/5]$ stabilized using either DMRG or AFQMC~\cite{Zhang1997-ConstrainedPathMonte}, in the $6$, $8$, or $12$ leg Fermi-Hubbard model at moderate to high doping $\delta = 1/8$ or $\delta = 1/5$.
In particular, this confirms the earlier finding of \cite{Jiang2021-GroundstatePhaseDiagram}, which reports $\nu = 2/3$ in the $6$ leg model at $\delta \ge 1/9$.
Recently, large scale DMRG simulations of the $8$ leg model typically find $\nu = 1/2$ at $t' \le 0$ and $\delta \ge 1/12$~\cite{Jiang2025-CompetitionChargedensitywaveSuperconductinga}.
We note that this work also observes a charge density wave without a commensurate spin density wave at large negative values of $t'$, and, depending on the boundary conditions, even on the electron doped side.
For the $t$-$t'$-$J$ model, \cite{Shen2024-GroundStateElectrondopeda} reports $\nu = 1/3$ stripes ($6$ leg) or uniform density ($8$ leg) at low doping $\delta = 1/12$ and fully filled stripes at higher doping.
Other works also show intermediate filling fractions $\nu = 1/2$~\cite{Jiang2021-GroundstatePhaseDiagram} and $\nu = 2/3$~\cite{Lu2024-EmergentSuperconductivityCompeting}. \\
\indent For the vanilla models at $t' = 0$,~\cite{Qin2020-AbsenceSuperconductivityPureb} discusses the extremely small difference in energy between the $\nu = 1$ ground state and the $\nu = 2/3$ stripes stabilized in~\cite{Jiang2024-GroundstatePhaseDiagram} in the $6$ leg Fermi-Hubbard system.
In the $8$-leg system, \cite{Xu2022-StripesSpindensityWaves} also finds the $\nu = 2/3$ and $\nu = 1$ stripe states to be indistinguishable in energy, using the CP-AFQMC method. By finite-size extrapolation, they expect $\nu = 1$ in the thermodynamic limit.
In one of the most comprehensive works on the $1/8$-doped vanilla Fermi-Hubbard model -- comparing the DMRG, AFQMC, iPEPS~\cite{Verstraete2004-RenormalizationAlgorithmsQuantumMany}, and DMET~\cite{Knizia2012-DensityMatrixEmbedding} methods in systems up to $L_y = 12$ or impurity clusters of up to $18$-sites -- B.-X. Zheng et al.~\cite{Zheng2017-StripeOrderUnderdoped} also find $\nu = 1$ to be the ground-state in the thermodynamic limit.
However, they find a broad range of stripe periods, $\lambda = 5$ to $\lambda = 8$, to be near-degenerate, corresponding to $\nu = 5/8$ to $\nu = 1$. We indicate this range in Fig.~\ref{fig_filling_t_prime_diagram}.
Meanwhile, the $6$ leg $t$-$J$ cylinder features $\nu = 2/3$ stripes at low doping~\cite{Lu2024-EmergentSuperconductivityCompeting}, while~\cite{Jiang2021-GroundstatePhaseDiagram} shows half-filled stripes in the $8$ leg case.\\
\indent Finally, the electron-doped side features uniform-density superconducting or antiferromagnetic phases at all but high doping in both the Fermi-Hubbard and $t$-$t'$-$J$ models~\cite{Jiang2021-GroundstatePhaseDiagram, Jiang2024-GroundstatePhaseDiagram, Lu2024-EmergentSuperconductivityCompeting}.
At high doping $\delta \gtrsim 0.2$,~\cite{Jiang2021-GroundstatePhaseDiagram} identifies stripe order with the period $\lambda$ commensurate with fully filled stripes. \\
\indent Very recently, variational methods have started to reach very large system sizes of up to $16 \times 16$ sites~\cite{Liu2025-AccurateSimulationHubbarda, Gu2025-SolvingHubbardModel, Roth2025-SuperconductivityTwodimensionalHubbarda}.
In clusters with fully periodic boundary conditions, these also report partially filled stripes with the commensurate filling $\nu = 1/2$ at $t' < 0$, while filled stripes ($\nu=1$) are favored at $t'=0$~\cite{Gu2025-SolvingHubbardModel, Roth2025-SuperconductivityTwodimensionalHubbarda}.
\begin{figure}[t!!]
\centering
\includegraphics[width=\linewidth]{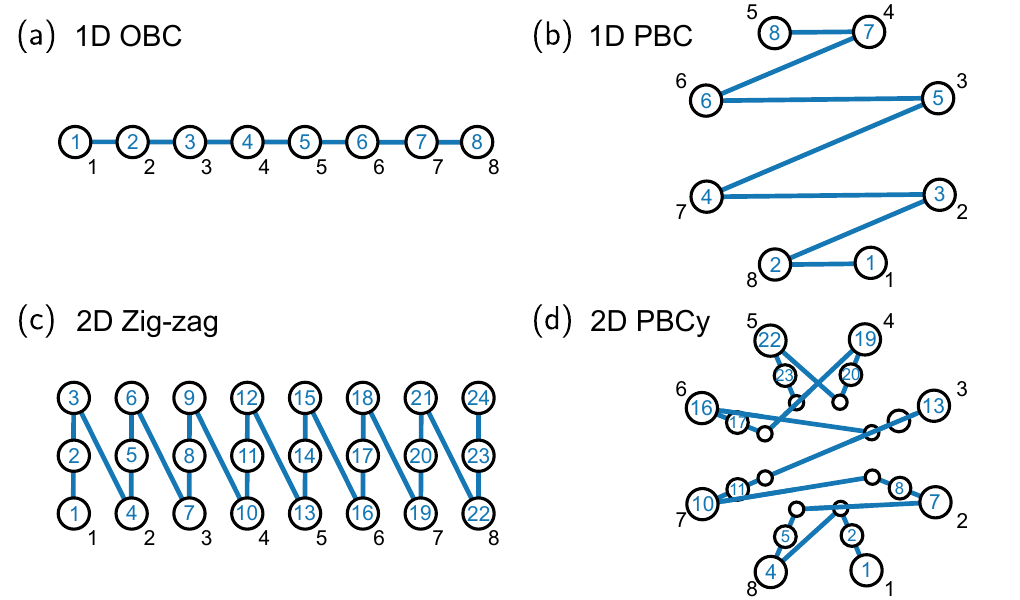}
\caption{
{MPS Mapping:}~Different ways to path the one-dimensional MPS (blue line) through the system's sites (black circles) are advantageous depending on the system's dimensionality and boundary conditions.
Blue numbers index the sites in the MPS, black numbers label the coordinate $x$ (a-c), or $y$ (d).
In our work, we use mapping (d) which allows us to study short, wide cylinders.
(a) Mapping typical for one-dimensional systems. This mapping is optimal for OBCs, where the physical interaction range is the same as that of the MPS. Periodic boundary conditions would introduce a (physical) nearest-neighbor bond of MPS range $L - 1$ making this mapping unsuitable for PBCs.
(b) MPS mapping for one-dimensional periodic systems~\cite{Wilke2023-SymmetryprotectedBoseEinsteinCondensation}. Compared to (a), the MPS range of most nearest-neighbor bonds is doubled to $2$. Even with PBCs, there are no nearest-neighbor bonds with MPS ranges greater than $2$, making calculations possible.
(c) Zig-zag mapping typical for two-dimensional systems. Nearest-neighbor interactions in $y$-direction are also nearest-neighbor in the MPS. NN interactions in the $x$-direction have MPS range $L_y$ making calculations exponentially costly in the cylinder width. PBCs in the $y$-direction add bonds of range $L_y - 1$, so cylindrical boundary conditions add little complexity compared to open ones. The combination of these features makes cylinders with $L_y < L_x$ a popular choice for DMRG studies.
(d) Combining (b) and (c) leads to an MPS mapping suited for two-dimensional systems with fully periodic boundary conditions or wide cylinders.
Comparable to (c), nearest-neighbor bonds in the $x$-direction have MPS range 1, while most nearest-neighbor bonds in $y$-direction have range $2 \: L_x$ -- combining (c) with the doubling encountered in (b).
This further restricts $L_x$ but enables studying two-dimensional systems with a long periodic direction using MPS.
}
\label{supp_fig_mps_mapping}
\end{figure}
As a prominent example of stripes in cuprate superconductors, we also indicate the approximately half-filled stripes reported in LSCO near $1/8$-doping in Fig.~\ref{fig_filling_t_prime_diagram}.
These were first suspected based on inelastic neutron scattering in $\mathrm{La}_{2-x}\mathrm{Sr}_x\mathrm{CuO}_4$~\cite{Cheong1991-IncommensurateMagneticFluctuations} and later confirmed by neutron diffraction in $\mathrm{La}_{1.6-2x}\mathrm{Nd}_{0.4}\mathrm{Sr}_x\mathrm{CuO}_4$ where the stripe pattern is pinned~\cite{Tranquada1995-EvidenceStripeCorrelations}.
\section{MPS Mapping}
\label{suppsec:MPS}
To facilitate the unconventional system geometry, featuring a long periodic direction $L_y > L_x$, we design a tailored mapping of the one-dimensional MPS through our two-dimensional system. 
The motivation behind this mapping is outlined in Fig.~\ref{supp_fig_mps_mapping}, which assembles the mapping from the traditional `zig-zag' mapping for two-dimensional systems, as well as a mapping for one-dimensional systems with periodic boundary conditions~\cite{Wilke2023-SymmetryprotectedBoseEinsteinCondensation}.
To estimate the characteristics of a given mapping, it is useful to think of the `MPS range' of interactions, i.e., the distance measured along the path followed by the MPS.
Based on this criterion, Fig.~\ref{supp_fig_mps_mapping} discusses the expected challenges and computational costs for different system geometries.
\\
\indent The four different mappings from the system coordinate(s) $x$ (1D) or $(x, y)$ (2D) to the MPS index $i$, are given as follows:
\begin{align}
    i^\mathrm{1D}_\mathrm{OBC}(x) &= x \ , \\
    i^\mathrm{1D}_\mathrm{PBC}(x) &=
    \begin{cases}
        \: 2x - 1 \quad &\mathrm{if} \ x \le L/2 \\
        \: 2(L-x+1) \ &\mathrm{if} \ x > L/2
    \end{cases}
    \quad , \\
    i^\mathrm{2D}_\mathrm{zig \myphen zag}(x, y) &= (x-1) \cdot L_y + y \ , \\
    i^\mathrm{2D}_\mathrm{PBCy}(x, y) &=
    \begin{cases}
         2 \: L_x (y -1) + x &\mathrm{if} \ y \le L_y/2 \\
         2 \: L_x(L_y-y)+L_x+x &\mathrm{if} \ y > L_y/2
    \end{cases}
    \ .
\end{align}
\section{Squeezed-Space Prediction}
In the main text, we compare the cross-lattice spin correlations $C^{(L_x)}_{\hat{S}^z}$ to a prediction obtained from just the charge structure combined with the AFM correlations of the undoped system.
The approach is motivated by thinking of the stripe as a one-dimensional line of dopants traversing a two-dimensional system -- making it comparable to the problem of a single dopant in a one-dimensional system.
In this setting, the dopant can tunnel without frustrating the AFM background if the spins neighboring the dopant are anti-aligned, i.e., if an AFM spin domain wall forms across the dopant.
Therefore, our squeezed-spaced prediction $C^{(L_x)}_\mathrm{pred}$ is defined in a one-dimensional system of length $L_x$ as follows:
In the undoped AFM, the spin correlations are given by $c{(x)} = \langle \hat{S}^z_1 \hat{S}^z_x\rangle$
Upon doping, three cases are important to form the qualitative prediction outlined above: i) If there are no dopants, we predict $C^{(L_x)}_\mathrm{pred} = c{(L_x)}$. ii) If there is exactly one dopant in the center of the system, we expect the spin correlations to be shifted by one site. $C^{(L_x)}_\mathrm{pred} = c{(L_x-1)}$, which has the opposite sign ($\leftrightarrow$ domain wall). iii) If there are dopants on both central sites of the system, we expect $C^{(L_x)}_\mathrm{pred} = c{(L_x-2)}$, returning to the individual sign of correlations.
In cases outside (i-iii), there is a dopant sitting on the edge of the system, meaning that the spin correlations from edge to edge vanish.
We can evaluate the probabilities for cases (i-iii) via the dopant densities $\langle \hat{n}_i \rangle$ and correlation functions $\langle \hat{n}_i \hat{n}_j \rangle$, $\langle \hat{n}_i \hat{n}_j \hat{n}_k \rangle$, $\langle \hat{n}_1 \hat{n}_2 \hat{n}_3 \hat{n}_4 \rangle$, as
\begin{align}
    p_0 &= \langle (1 - \hat{n}_1) (1 - \hat{n}_2) (1 - \hat{n}_3) (1 - \hat{n}_4) \rangle \ , \\
    p^\mathrm{(ctr)}_1 &= \langle (1 - \hat{n}_1) \: \hat{n}_2 \: (1 - \hat{n}_3) (1 - \hat{n}_4) \rangle \\
    &\quad \ + \langle (1 - \hat{n}_1) (1 - \hat{n}_2) \: \hat{n}_3 \: (1 - \hat{n}_4) \rangle\ , \ \mathrm{and}\\
    p^\mathrm{(ctr)}_2 &= \langle (1 - \hat{n}_1) \: \hat{n}_2  \: \hat{n}_3 \: (1 - \hat{n}_4) \rangle \ . \\
\end{align}
This gives the combined prediction
\begin{equation}
    C^{(L_x)}_\mathrm{pred} = p_0 \: c{(L_x)} + p_1^\mathrm{(ctr)} c(L_x - 1) + p_2^\mathrm{(ctr)} c(L_x - 2) \ ,
\end{equation}
displayed along the true spin correlations in Fig.~\ref{fig_szszs}.
In practice, we obtain the spin and charge correlations by averaging over all legs of the cylinder.
%
%
%
\section{DMRG}
We use DMRG in the framework of matrix product states (MPS)~\cite{Schollwoeck2011-DensitymatrixRenormalizationGroup} implemented in the SyTeN toolkit~\cite{hubig:_syten_toolk}.
We study cylindrical systems that feature periodic boundary conditions (PBCs) in one direction ($y$-direction) and open boundary conditions (OBCs) in the other ($x$-direction).
\begin{figure}[t!!]
\centering
\includegraphics[width=\linewidth]{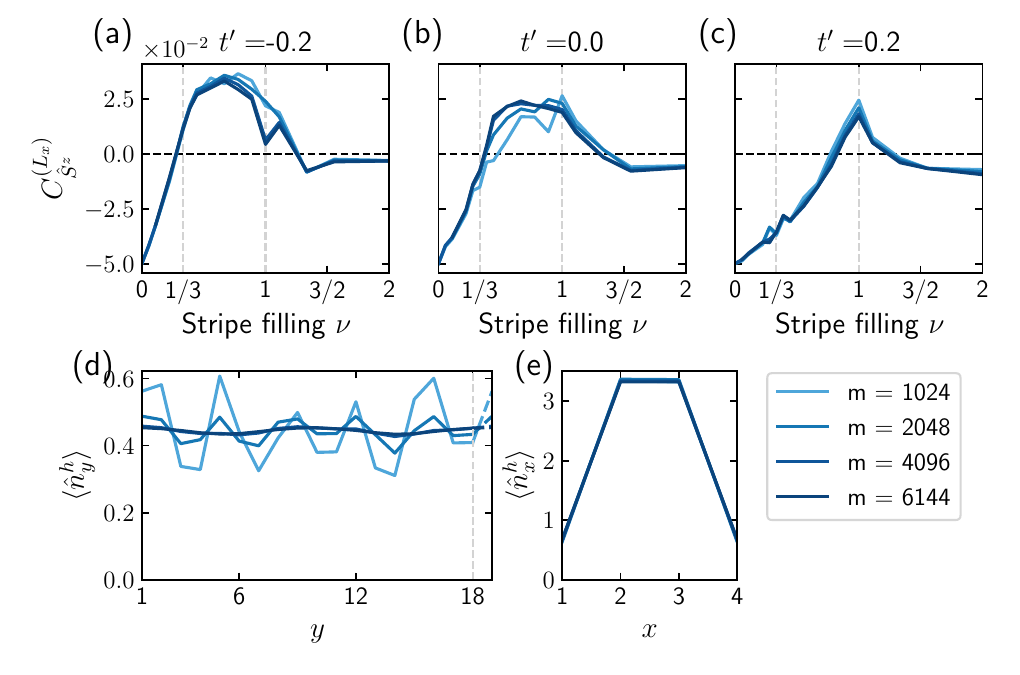}
\caption{
{DMRG Convergence:}
(a-c) Tracking the correlations $C^{(L_x)}$, presented in Figs.~\ref{fig_filling_t_prime_diagram} and \ref{fig_szszs}, with increasing bond dimension $m$.
The curves converge at $m \approx 4 000$.
(d, c) Dopant densities for parameters $t' = -0.2$ and $\nu = 4/9$, summed along the $x$ ($y$) direction, plotted along the $y$ ($x$) direction.
$\langle \hat{n}^h_y \rangle$ becomes flat at $m \approx 4 000$.
}
\label{supp_fig_convergence}
\end{figure}
\begin{figure*}[t!!]
\centering
\includegraphics[width=\textwidth]{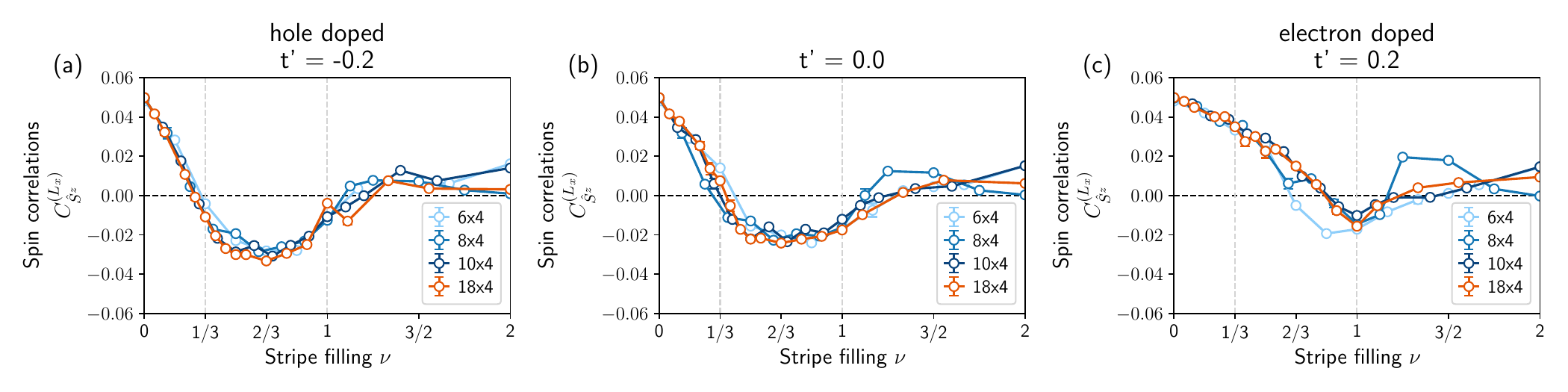}
\caption{
{Size Comparison:} Comparing the cross-lattice spin-spin correlations in a large $18$-leg system to smaller systems with $L_y \leq 10$, typically used to study the model's finite-doping phase diagrams.
(a-c) At all investigated values of $t'$, we find good qualitative agreement between different system sizes, 
highlighting the robustness of the single-stripe perspective.
}
\label{supp_fig_size_comparison}
\end{figure*}

While this type of geometry is typical for DMRG studies, the direction featuring PBCs is usually much shorter than the one featuring OBCs ($L_y \ll L_x$). In contrast, we study the case of very wide cylinders ($L_y \gg L_x$), facilitated by a tailored MPS mapping. \\
\indent We exploit the full $\mathrm{U}(1) \times \mathrm{SU}(2)$ symmetry of the particle number and spin rotation, working with a fixed particle number $N$ and total spin $S = 0$ ($N$ even) or $S = \frac{1}{2}$ ($N$ odd).
To find the ground state, we use a mixture of the single-site and two-site DMRG algorithms, starting from a Fermi-sea initial state.
To reduce the risk of getting stuck, we utilize global subspace expansion, as proposed for use in time evolution in Ref.~\cite{Yang2020-TimeDependentVariational}, to increase the bond dimension in the early stages of our calculations. \\
\indent The observables we investigate are well-converged using bond dimensions up to $m = 6 \: 144$,
and we find the effective $\mathrm{U}(1)$ bond dimension of $\mathrm{SU}(2)$-calculations with $m \approx 6 \, 000$ to be $m^\mathrm{eff.} \approx 18 \, 000$.
For the high-probability product states, presented in Fig.~\ref{fig_internal_structure}, we reduce the symmetry to $U(1) \times U(1)$ to facilitate snapshots in the $S^z$ basis. The calculations are performed with a maximum bond dimension of $m = 10 \: 240$.
%
\section{Convergence}
\label{suppsec:convergence}
Due to the unconventional mapping of the MPS, it is especially important to closely monitor the convergence of the DMRG calculations.
In Fig.~\ref{supp_fig_convergence}(a)-(c), we track the correlations $C^{(L_x)}$ with increasing bond-dimension $m$.
We find this important quantity, which forms the basis of our analysis, to be converged at $m \approx 4000$.
Complementary to the correlations, we also present convergence data for the dopant density in Fig.~\ref{supp_fig_convergence}(d,e).
The figures shows the summed densities
\begin{align}
    \langle \hat{n}^h_y \rangle
    = \langle \sum_{x} \hat{n}^h_{(x,y)} \rangle \ , \ \mathrm{and} \\
    \langle \hat{n}^h_x \rangle
    = \langle \sum_{y} \hat{n}^h_{(x,y)} \rangle \ ,
\end{align}
for the reference parameters $t' = 0.2$ and $\nu = 4/9$.
Even at the lowest bond dimension $m \approx 1000$, $\langle \hat{n}^h_x \rangle$ is nicely symmetric, and shows lower dopant densities on the edges of the system, compared to the bulk.
As the main challenge we face is the extended, periodic $y$-direction, $\langle \hat{n}^h_y \rangle$ is a particularly important probe for convergence.
At low bond dimensions $m \lesssim 2000$, the density is peaked at regular intervals, reminiscent of the pairing structure presented in Fig.~\ref{fig_internal_structure}(a). At $m \approx 4000$, the density becomes flat, reflecting the periodic boundary conditions.
From this we conclude that the algorithm is able to capture the long periodic direction, using the MPS mapping tailored for this situation.
\section{System Size Comparison}
\label{suppsec:size_comparison}
To strengthen the connection between our single-stripe study and finite-doping phase diagrams, compare our $18$-leg results to isolated stripes in $4 \times 6$, $4 \times 8$, and $4 \times 10$ systems, bridging the gap to cylinder widths commonly used in the literature.
In Fig.~\ref{supp_fig_size_comparison}, we compare the $C^{(L_x)}_{\hat{S}^z}$ correlations between these systems.
While there are notable deviations on the electron-doped side for the $6$-leg system, we find good overall agreement, especially for the $L_y \geq 8$ cases.
This suggests that the physics of isolated stripes is already captured well in these small systems -- an import result supporting the validity of contemporary numerical studies.
\begin{figure*}[t!!]
\centering
\includegraphics[width=\textwidth]{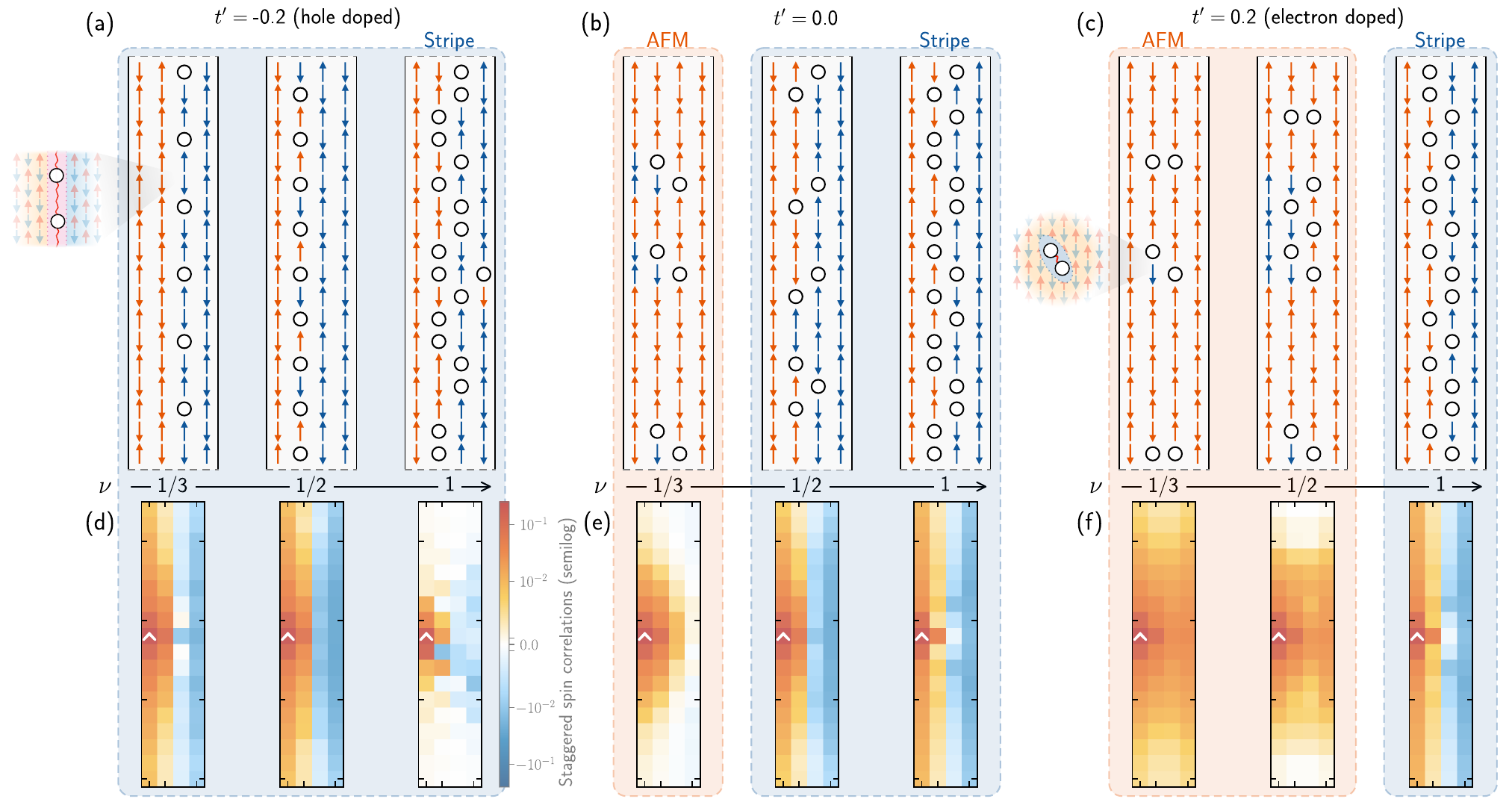}
\caption{
{High-Probability Product States and Spin Correlations:} (a-c) Comparing the stripes' microscopic structure across different filling fractions $\nu$ and different values of $t'$. The central ($\nu = 1/2$) column of each subfigure is the same as presented in Fig.~\ref{fig_internal_structure} of the main text.
(a) On the hole-doped side ($t' = 0$), a stripe is formed at fillings as low as $\nu = 1/3$. The dopants are separate and form a line around the cylinder. The separation decreases with increasing filling.
(b) In the vanilla model ($t' = 0$), the dopants form tightly bound, next-nearest-neighbor pairs. However, a collective stripe instability still begins to form between $\nu = 1/3$ and $1/2$.
This behavior reflects the near-degeneracy between two distinct pair configurations identified in a previous study~\cite{Blatz2025-TwoDopantOriginCompetinga}.
(c) On the electron-doped side ($t' = 0.2$), we also observe tightly-bound pairs, but a spin domain wall only begins to form near $\nu = 1$.
The tendency of individual pairs to form a stripe is suppressed.
Across all values of $t'$, the microscopic structure at $\nu = 1$ is very similar. This is in line with our interpretation in terms of a squeezed space picture, which is independent of the underlying pairing properties.
(d-e) Staggered spin-spin correlation functions with respect to the reference spins marked by white hat symbols.
The emergence of a spin domain wall in the spin correlations coincides with its appearance in the high-probability product states.
}
\label{supp_fig_microscopic_structure}
\end{figure*}
Furthermore, this finding is in line with a few-dopant origin of the stripe physics, such as the single-pair picture we propose.
\section{High-Probability Product States}
In addition to the data presented in Fig.~\ref{fig_internal_structure} of the main text, we present high-probability product states for multiple filling fractions $\nu$ in Fig.~\ref{supp_fig_microscopic_structure}(a)-(c).
By comparing to the straggered spin-spin correlation functions presented in Fig.~\ref{supp_fig_microscopic_structure}(d)-(f), we verify that the high-probability product states correctly capture the formation of a spin domain wall.
The comparison between different $\nu$'s highlights the trend of a spin domain wall emerging at lower filling on the hole-doped side, compared to the electron-doped side.
The microscopic structure uncovered by the product states exemplifies the physical picture we propose.
At low filling fractions, the high-probability product states feature separate holes at $t' < 0$, while the dopants form tightly-bound pairs on the electron-doped side ($t' > 0$).
This is in line with the structure of a single pair, investigated in a previous work~\cite{Blatz2025-TwoDopantOriginCompetinga}.
Meanwhile, the $t'$-dependence of the fully filled stripe ($\nu = 1$) is minimal, which is expected from the squeezed-space picture.
\end{document}